\documentclass[reprint,amsmath,amssymb,aps,prb,showpacs,floatfix]{revtex4-1}
\usepackage{graphicx}
\usepackage{epstopdf}
\DeclareGraphicsExtensions{.eps, .jpg, .png}
\usepackage{dcolumn}

\begin{document}

\title{Carbon Nitride Frameworks and Dense Crystalline Polymorphs}

\author{Chris J.\ Pickard} \email{cjp20@cam.ac.uk}
\affiliation{Department of Physics \& Astronomy, University College
  London, Gower Street, London WC1E 6BT, UK, and Department of
  Materials Science $\&$ Metallurgy, University of Cambridge, 27
  Charles Babbage Road, Cambridge CB3 0FS, UK}

\author{Ashkan Salamat} \affiliation{Lyman Laboratory of Physics,
  Harvard University, Cambridge, MA 02138, USA, and Department of
  Physics and Astronomy and HiPSEC, University of Nevada Las Vegas,
  Las Vegas, Nevada 89154, USA}

\author{Michael J.\ Bojdys} \affiliation{Charles University in Prague
  Faculty of Science, Hlavova 8, 128 43 Praha 2, Czech Republic, and
  Institute of Organic Chemistry and Biochemistry ASCR, v.v.i.
  Flemingovo n\'{a}m. 2, CZ-166 10 Prague 6, Czech Republic, and Head
  of the Functional Nanomaterials Group Joint laboratory of IOCB and
  Charles University in Prague}

\author{Richard J.\ Needs} \affiliation{Theory of Condensed Matter
  Group, Cavendish Laboratory, J J Thomson Avenue, Cambridge CB3 0HE,
  UK}

\author{Paul F.\ McMillan} \affiliation{Department of Chemistry,
  Christopher Ingold Laboratory, University College London, Gordon
  Street, London WC1H 0AJ, UK}

\date{\today}

\begin{abstract}
  We used \textit{ab initio} random structure searching (AIRSS) to
  investigate polymorphism in C$_3$N$_4$ carbon nitride as a function
  of pressure.  Our calculations reveal new framework structures,
  including a particularly stable chiral polymorph of space group
  $P4_32_12$ containing mixed $sp^2$ and $sp^3$-bonding, that we have
  produced experimentally and recovered to ambient conditions.  As
  pressure is increased a sequence of structures with fully
  $sp^3$-bonded C atoms and three-fold coordinated N atoms is
  predicted, culminating in a dense $Pnma$ phase above 250 GPa. Beyond
  650 GPa we find that C$_3$N$_4$ becomes unstable to decomposition
  into diamond and pyrite-structured CN$_2$.
\end{abstract}

\maketitle

\section{Introduction}

Carbon nitrides with N:C ratios $>$1 form a class of solid state
compounds with properties that are being investigated for applications
ranging from gas sorption and catalysis to energy conversion and
storage
\cite{Kroke_Novel_group_14_nitrides_2013,Schwarzer_Tri-s-triazines_2013,Thomas_graphitic_carbon_nitride_2008,Goglio_carbon_nitride_synthesis_2008,Malkow_CN_2000,Wang_photocatalyst_2009}.
Experimentally known structures are based on $sp^2$-bonded carbon
atoms and include polymeric to graphitic layered compounds. Within the
C-N-H system, materials related to Liebig's melon
((C$_6$N$_9$H$_3$)$_n$) contain tri-$s$-triazine (heptazine:
C$_6$N$_7$) structural units linked $via$ -NH- groups
\cite{Kroke_Novel_group_14_nitrides_2013,Schwarzer_Tri-s-triazines_2013,Lotsch_melon_2007,Jurgens_melem_2003,Lotsch_melam_2007}.
Additional graphitic materials containing intercalated Li$^+$, Cl$^-$
and Br$^-$ ions are formed by polytriazine imide (PTI) layers
consisting of triazine (C$_3$N$_3$) rings linked by -N= or -NH-
species
\cite{Bojdys_synthesis_crystalline_gCN_2008,Wirnhier_Intercalation_2011,Chong_tuning_2D_CN_2013,Zhang_C3N4_derivative_2001}.
Synthesis of macroscopic flakes of triazine-based graphitic carbon
nitride (TGCN) composed of nitrogen-linked triazine (C$_3$N$_3$) units
was reported recently \cite{Algara-Siller_triazine_g-CN_2014}. This
structure was first proposed in the mid-1990's as ``graphitic carbon
nitride'' (``g-C$_3$N$_4$'') by analogy with the structurally related
graphite
\cite{Liu_low-compressibility_1994,Teter_low_compressibility_1996}. A
nanocrystalline material with C$_3$N$_4$ stoichiometry had been
reported from chemical vapor deposition
\cite{Todd_synthesis_CN_1995,Kouvetakis_synthesis_1994} and related
models were studied theoretically
\cite{Ortega_hexagonal-planar_1995,Lowther_g-CN_1999,Teter_low_compressibility_1996}. Triazine-based
carbon nitrides can also be exfoliated to produce layered materials
analogous to few-layered graphene
\cite{Bojdys_2D_CN_2013,Ma_2D_g-CN_2014}.  Theory shows that the
optical bandgap (1.6--2.7 eV) important in photocatalysis can be tuned
by controlling the layer buckling at high pressure
\cite{Deifallah_2D_CN_2008,Zuluaga_band_gap_g-C3N4_2015}.

Interest in dense carbon nitride polymorphs based on $sp^3$ bonded C
atoms began with predictions from density functional theory (DFT)
calculations of superhard properties
\cite{Cohen_calculation_bulk_moduli_1985,Liu_low-compressibility_1989,Liu_structure_low-compressibility_1990,Sung_CN_superhard_1996,Teter_low_compressibility_1996}.
First investigations began by considering known structures such as
those of Si$_3$N$_4$ ceramic phases, and more recently using structure
searching methods \cite{Dong_carbon_nitrides_2015}. However none of
the predicted high density C$_3$N$_4$ polymorphs has been demonstrated
experimentally to date
\cite{Kroke_Novel_group_14_nitrides_2013,Kroke_Angewandte_Highlights_gt-C3N4}.
One carbon nitride solid containing $sp^3$ bonded C atoms was produced
by laser heating dicyandiamide (DCDA: C$_2$N$_4$H$_4$) at $P>27$ GPa
in a diamond anvil cell
\cite{Horvath-Bordon_high-pressure_CN_2007}. This material with a
defective wurtzite structure could be recovered to ambient conditions
\cite{Salamat_tetrahedral_CN_2009}. A recent study has identified
formation of a CN phase that also contains $sp^3$ bonded carbon
\cite{Goncharov_CN}.

Experimental compression studies and DFT calculations have also
revealed the existence of carbon nitride structures containing both
$sp^2$ and $sp^3$ bonded C atoms obtained from the graphitic layered
PTI compound C$_6$N$_9$H$_3$$\cdot$HCl. Layer buckling causes C-N
distances in adjacent layers to approach a covalently bonded value and
an interlayer bonded (ILB) form becomes stabilized above 47 GPa
\cite{Salamat_pillared-layered_CN_2013}. This result shows that carbon
nitride framework structures containing mixed $sp^2$ and $sp^3$
bonding can exist over a range of densities. Microporous carbon
nitride solids have also been produced at ambient pressure by linking
C$_3$N$_3$ triazine units into three-dimensional covalent networks
\cite{Vodak_C3N4_networks_2003,Ren_triazine_2012,Bojdys_triazine_2010,Li_porous_g-C3N4_2015,Reece_triazine-based_framework_2015}.

The structural polymorphism of $sp^3$-bonded carbon nitride is
usefully compared with that of elemental carbon. Here high-pressure
synthesis is well known to produce dense structures
\cite{Oganov_carbon_2013}. Low pressure allotropes dominated by $sp^2$
bonding consist of planar graphite and graphene structures formed from
fused six-membered rings. Introducing curvature into the layers leads
to fullerenes and C-nanotubes, while hierarchical low density
mesoporous solids assembled at different length scales are important
for nanotechnology and planetary science studies
\cite{Oganov_carbon_2013,Gong_design_fabrication_2014,Lazzeri_carbon_nanoscience_2014}.
Theoretical modelling now also points to the potential existence of
metastable zeolitic $sp^3$-bonded carbon framework structures in the
0--20 GPa range \cite{Baburin_zeolite_carbon_2015}.  In equilibrium at
low temperatures and pressures above 0.76 GPa
\cite{Bundy_1961,Dia-graphite_mechanism_Parrinello_2011} the dense
$sp^3$-bonded structure of diamond becomes the most stable form and no
further structural transformations are expected until a body-centered
(BC8) polymorph is predicted above 1000 GPa (1 TPa)
\cite{Oganov_carbon_2013}.  Other new dense allotropes relevant to
understanding C-rich exoplanets are predicted to exist at extreme
pressures extending into the TPa range
\cite{Martinez-Canales_multiterapascal_carbon_2012,Smith_5TPa_carbon_2014,Pickard_piling_up_pressure_2014}.

Here we used \textit{ab initio} random structure searching (AIRSS)
\cite{Pickard_silane_2006,AIRSS_2011} to explore carbon nitride
polymorphism as a function of density throughout a wide pressure range
extending between 0--1000 GPa (0--1 TPa), focusing on behavior at the
C$_3$N$_4$ composition.  At low density the results reveal new
crystalline architectures including previously unsuspected framework
structures based on mixed $sp^2$-$sp^3$ bonding.  One of these
structures ($P4_32_12$ or its enantiomorph $P4_12_12$) has been
verified experimentally in our study. The AIRSS results allow a
ranking of stability among predicted C$_3$N$_4$ polymorphs up to 650
GPa, above which C$_3$N$_4$ becomes unstable relative to diamond +
pyrite-structured CN$_2$.

\section{Structure searching}

AIRSS is a powerful theoretical prediction technique that has led to
the discovery of new structures subsequently verified by experiment,
in materials ranging from hydrogen
\cite{Pickard_hydrogen_2007,Pickard_phase_IV_hydrogen_2012}, ammonia
\cite{Ammonia_Ninet_2014}, ammonia hydrates
\cite{Fortes_AMH-II_2009,Griffiths_AD-II_2012}, aluminium hydride
\cite{Pickard_AlH3_2007}, silane \cite{Pickard_silane_2006} to xenon
oxides \cite{Xenon_oxides_2015}.  Within the AIRSS method a cell
volume and shape are selected at random, atoms are added at random
positions to provide the desired composition, and the system is
relaxed until the forces on the atoms are negligible and the pressure
takes the required value. This procedure is repeated many times,
giving a relatively sparse sampling of the ``structure space''. We
typically constrain the minimum separations between atom pairs (C--C,
C--N, N--N) in the initial structures, which helps to space out the
atoms appropriately, while retaining a high degree of randomness.
These minimum distances are obtained from preliminary short AIRSS
runs.  In almost all of the subsequent searches we impose symmetry
constraints on the structures, for example, we might demand that all
structures belong to randomly selected space groups with 8 or more
symmetries.  In the next stage of searching we might choose to allow
structures with 4 or more symmetries, etc.  With this approach we
obtain a very good sampling of the high-symmetry structures, which can
be extended towards lower symmetries as the available computing
resources allow.  The use of symmetry constraints greatly reduces the
size of the ``structure space'', although they also break it up into
disconnected regions which can prevent some structures from relaxing
fully.  However, there is a strong tendency for low energy structures
to possess symmetry \cite{Pauling_1929}, and we find that the
application of symmetry constraints throughout the searching is a very
useful tool.  Searches for low pressure structures are
performed using heptazine- and triazine-based units and we consider
structures containing up to sixteen formula units (f.u.).

The structure searches were performed using AIRSS and first-principles
DFT methods, version 8.0 of the \textsc{castep} plane-wave basis set
pseudopotential code \cite{ClarkSPHPRP05}, default \textsc{castep}
ultrasoft pseudopotentials\cite{Vanderbilt90}, a plane-wave basis set energy cutoff of 280
eV, and an initial Brillouin zone sampling grid of spacing
2$\pi\times$0.1\ \AA$^{-1}$.  The PBE generalized gradient
approximation density functional \cite{PBE_1996} was used for the
structure searches.  The final numerical results reported in this
paper were obtained using the PBEsol functional with a basis set
energy cutoff of 700 eV,  \textsc{castep} ``on-the-fly'' ultrasoft
pseudopotentials, as described in the Supplemental
Material\cite{Supplemental}, 
and a $k$-point spacing of 2$\pi\times$0.05\
\AA$^{-1}$, except for the data in Table I, which were calculated with
a finer $k$-point spacing of 2$\pi\times$0.03\ \AA$^{-1}$.

%Searching: kpoint sampling 0.1 x 2pi and 280 eV using the old file
%usps {C,N}_00.usp
%
%Final: 0.05 x 2pi and 700 eV (grid scale 2, fine grid scale 2.3) using
%CASTEP 8.0 OTFG USPs

\section{Choice of density functional}

Estimating the pressure of polymorphic transitions among $sp^2$ and
$sp^3$ bonded phases for carbon nitride materials is difficult because
no reliable data for the structures and energetics of $sp^3$ bonded
C$_3$N$_4$ polymorphs are available.  In order to choose a density
functional that would provide reasonable transition pressures for
comparison with experiment we considered the well-known transition
from $sp^2$-bonded graphite to $sp^3$-bonded diamond in elemental
carbon, which has distinct similarities to C$_3$N$_4$.  The structures
and lattice constants of diamond and graphite are known, which allows
accurate comparisons to be made with the results of DFT calculations
\cite{Bundy_1961,Dia-graphite_mechanism_Parrinello_2011}.

\begin{figure}
  \includegraphics[width=7cm]{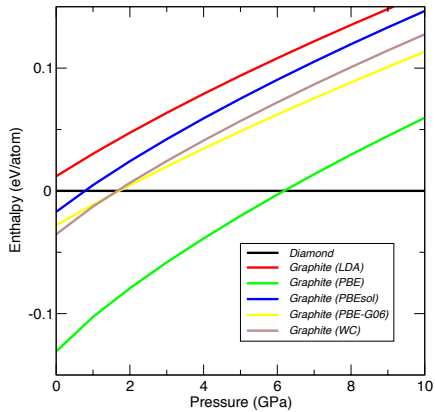}
  \caption{(Color online) Enthalpy per atom for carbon graphite
    relative to diamond as a function of pressure calculated using
    five different density functionals (LDA, PBE, PBEsol, PBE-G06 and
    WC).}
  \label{fig:enthalpy_dia_graph}
\end{figure}

In Fig.\ \ref{fig:enthalpy_dia_graph} we plot the differences in
enthalpy between the graphite and reference diamond structures
calculated using five density functionals: the local density
approximation (LDA) \cite{Kohn_LDA_1965}, the semi-local PBE
\cite{PBE_1996}, WC \cite{Wu_Cohen_functional_2006} and PBEsol
generalized gradient approximations \cite{PBEsol_2008}, and the
PBE+G06 generalized gradient approximation with empirical van der
Waals corrections \cite{PBE_1996,Grimme_G06}.  The calculated
graphite/diamond transition pressures ranged from -1 GPa (LDA) to +6
GPa (PBE), which provides an estimate of the range of errors in the
pressure arising from the approximate density functionals of about
$\pm$4 GPa, see Fig.\ \ref{fig:enthalpy_dia_graph}.

The most important factor for obtaining an accurate graphite/diamond
transition pressure is to use a functional that provides a good
account of the volume of diamond and the in-plane lattice constant of
graphite. Here the effects of zero-point vibrational motion are
estimated to be small and have been neglected. The LDA normally tends
to underestimate volumes, while PBE tends to overestimate them.
PBEsol, WC, and PBE+G06 give volumes for diamond intermediate between
those of LDA and PBE and are in reasonable agreement with experiment.
The description of graphite is more variable. The graphite/diamond
transition pressure is calculated to be 0.7 GPa using PBEsol, in close
agreement with experiment. WC and PBE-G06 both give 2 GPa, while PBE
predicts the transition at 6 GPa. We therefore elected to use PBEsol
for our studies of C$_3$N$_4$ polymorphs. The PBE and PBEsol
functionals gave the same sequence of polymorphic phase transitions
although the PBEsol transition pressures are somewhat larger.

\section{Structures found in the high pressure searches}

Enthalpy-pressure relations for C$_3$N$_4$ polymorphs found by AIRSS
using the PBEsol functional are shown in Fig.\
\ref{fig:enthalpy_C3N4_PBE_0-400_GPa}. Below approximately 5 GPa we
find that carbon nitride compounds are metastable with respect to
elemental carbon (diamond or graphitic polymorphs) and N$_2$
phases. At low pressure a wide variety of corrugated layers or open
framework structures based on linked heptazine or triazine motifs are
predicted to occur and the number of potential polymorphs and
structure types increases dramatically with system size
\cite{Gracia_corrugated_2009}.  These low-density materials typically
have very large unit cell volumes and they rapidly become destabilized
with increasing pressure \cite{Dong_carbon_nitrides_2015}.  The
importance of such mesoporous phases and their relationship to layered
and framework-structured carbon nitrides is discussed below.
Heptazine-based structures of $Cc$, $Fdd2$ and $Pbca$ symmetries are
the most stable below 2 GPa in PBEsol, and the $P4_32_12$ structure
with mixed $sp^2$-$sp^3$ bonding is stable in the range 2--11 GPa. At
higher pressures we find structures with $sp^3$ bonding around the
carbon atoms to be the most stable.  We found the following sequence
of transitions among the most stable phases as a function of pressure:
\begin{multline*}
Cc
\xrightarrow      {0.05}   Pbca
\xrightarrow      {2}      P4_32_12
\xrightarrow     {11}      P31c
\xrightarrow     {47}      Pnnm
\xrightarrow    {100}\\    Pmc2_1
\xrightarrow    {106}      I\bar{4}2m
\xrightarrow    {172}      I\bar{4}3d
\xrightarrow    {195}      Cmc2_1
\xrightarrow    {238}      Pnma
\end{multline*}
The numbers above the arrows indicate the PBEsol transition pressures
in GPa. (The $Cm$ structure found in DFT searches by Dong \textit{et
  al.}\ \cite{Dong_carbon_nitrides_2015} is not predicted to be stable
at any pressure because our study has discovered new structures that
are more stable.)

Structures of the high-density phases predicted to become stable at
high pressure are illustrated in Fig.\
\ref{fig:structures_high_pressure}.  The low-enthalpy structures
contain only C--N bonds, no C--C or N--N bonding was found to occur in
energetically competitive structures over the pressure range studied
of 0--1 TPa.  Between 11--47 GPa (using PBEsol) the
$\alpha$-C$_3$N$_4$ structure with $P31c$ symmetry analogous to the
ceramic phase $\alpha$-Si$_3$N$_4$ is predicted to be stable.  This is
one of the polymorphs considered by Cohen, Sung and others to
constitute a likely ultra-low compressibility, superhard material
\cite{Cohen_calculation_bulk_moduli_1985,Liu_low-compressibility_1989,Liu_structure_low-compressibility_1990,Sung_CN_superhard_1996,Teter_low_compressibility_1996}.
The $\alpha$-C$_3$N$_4$ structure contains $sp^3$ bonded 4-coordinated
C atoms and $sp^2$ bonded three-fold coordinated N atoms
characteristic of other low-enthalpy structures at high pressures. In
addition to its low compressibility and potentially high hardness it
has been predicted to have the interesting materials property of a
negative Poisson's ratio \cite{Guo_1995}.  The $\beta$-C$_3$N$_4$
phase considered in early discussions is not stable at any pressure
within the range studied.  Above 47 GPa we find that
$\alpha$-C$_3$N$_4$ is overtaken in stability by a $Pnnm$ structured
phase that remains the most stable polymorph until 100 GPa (Fig.\
\ref{fig:structures_high_pressure}). At this pressure a structure with
$Pmc2_1$ symmetry makes a brief appearance, before the introduction of
a cubic ($I\bar{4}2m$) structure \cite{Dong_carbon_nitrides_2015}
containing nearly tetrahedrally bonded N atoms along with the
$sp^3$-bonded C atom centers. Neither the $Pnnm$ nor $Pmc2_1$
polymorphs were identified as stable phases in structure searches
reported by Dong \textit{et al.}\ \cite{Dong_carbon_nitrides_2015}. At
195 GPa another cubic ($I\bar{4}3d$) phase makes a brief appearance, followed
by a $Cmc2_1$ structured polymorph that is stable until 238 GPa. This
phase was described by Dong \textit{et al.}\ as occurring between
224--300 GPa. It appears to resolve the competing structural situation
between C- and N-based bonding with the emergence of clearly separated
C- and N-atomic layers within a structure that could be described as
``defective'' diamond or c-BN. Above 238 GPa the most stable predicted
polymorph has orthorhombic $Pnma$ symmetry. This remains the most
stable phase until 650 GPa where it decomposes into diamond and a
pyrite-structured CN$_2$ polymorph according to
\begin{equation*}
  {\rm C}_3{\rm N}_4(Pnma) \rightarrow {\rm C}({\rm
    diamond}) + 2{\rm C}{\rm N}_2(Pa\bar{3}, {\rm pyrite}) \; .
\end{equation*}

%In the absence of such decomposition the $Pnma$ structure remains the
%most stable polymorph of C$_3$N$_4$ up to the highest pressure
%investigated (1 TPa).

\section{Diamond anvil cell experiments}

\begin{figure*}
  \centering
  \includegraphics[width=0.8\textwidth]{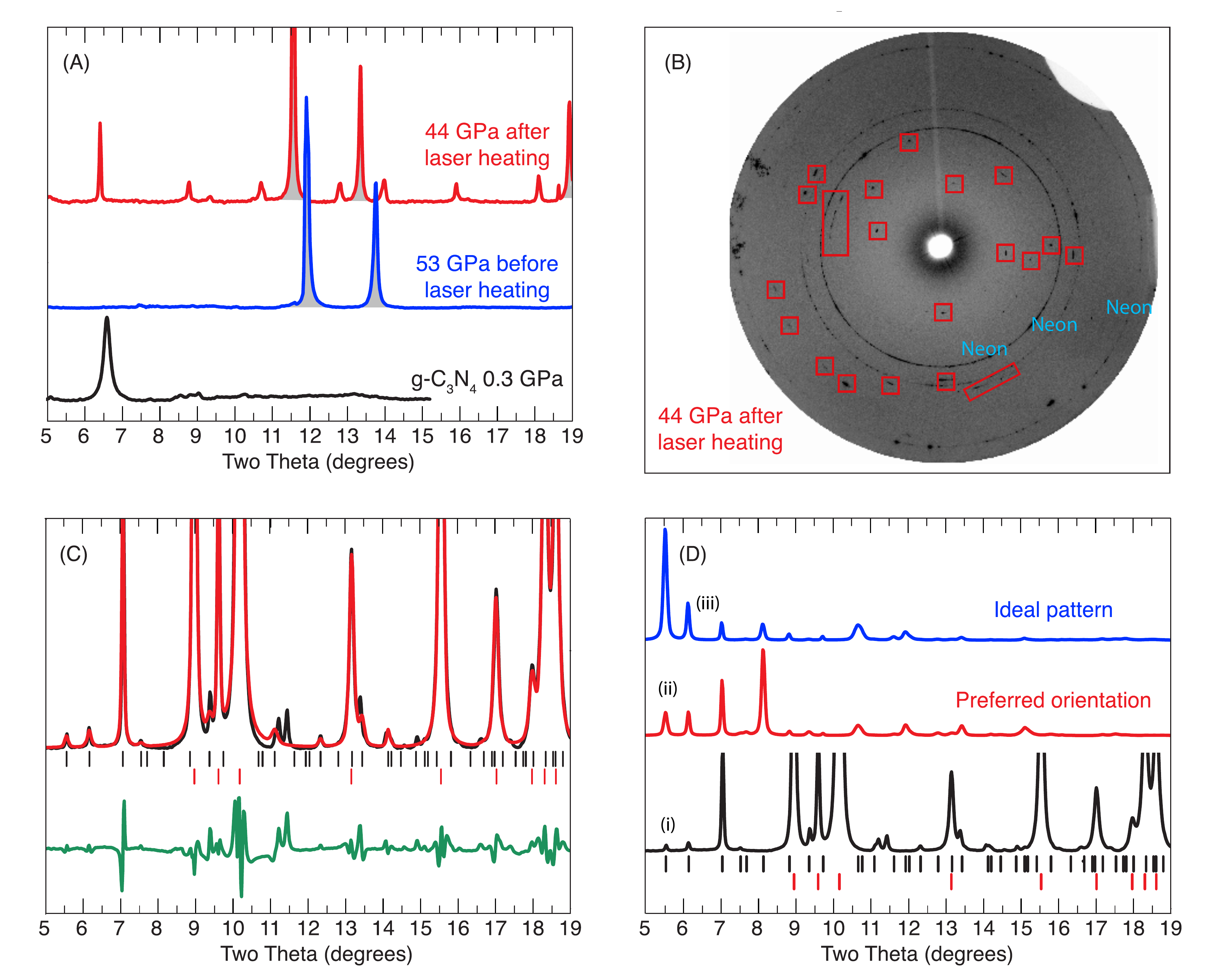}
  \caption{(Color online) Experimental X-ray diffraction data.  (A)
    Starting C$_3$N$_4$ sample before and after laser heating at
    various pressures. Peaks with shaded grey areas are from the Ne
    pressure transmitting medium.  (B) Diffraction image at 44 GPa
    after laser heating at 1500 K. The emergence of new crystalline
    features are highlighted in red boxes. The three Debye-Scherrer
    diffraction rings due to the Ne pressure transmitting medium are
    labelled.  (C) Le Bail refinement of the recovered $P4_32_12$ (or
    $P4_12_12$) phase of C$_3$N$_4$. Data points and Le Bail fits are
    overlaid in black and red, respectively, and a difference plot is
    shown. The experimental pattern contains a contribution from the
    Re gasket material that remained present surrounding the sample.
    Top (black) and bottom (red) tick marks are for the C$_3$N$_4$
    phase and Re, respectively.  (D) (i) The plot shows the
    experimental data obtained at ambient conditions. (ii) The plot
    depicts a simulated pseudo-Rietveld refinement of the data
    assuming a composite of oriented crystallites with fixed atomic
    positions taken from the theoretical study (Table
    \ref{table:exp_theory}). (iii) The plot shows the calculated
    powder pattern expected from complete sampling of randomly
    oriented crystallites.}
  \label{fig:experimental_data}
\end{figure*}

To begin to test our theoretical predictions we carried out an
exploratory synthesis experiment in a laser-heated diamond anvil cell
(DAC) using a sample of crystalline TGCN g-C$_3$N$_4$ as a starting
material \cite{Algara-Siller_triazine_g-CN_2014}. This compound has
been shown to have $P6_3cm$ or $P\bar{6}m2$ symmetry corresponding to
ABC or AB stacking of the buckled graphitic layers. A thin sample
flake of approximately 60 $\times$ 60 micrometers in lateral
dimensions was loaded into the DAC chamber and isolated from the
diamond window using three small ruby spheres. The pressure was
determined by ruby fluorescence from a different ruby sphere placed
away from the sample. The sample was compressed using Ne as the
pressure-transmitting medium (PTM).  \textit{In situ} X-ray
diffraction experiments were carried out at beamline ID27 of the
European Synchrotron Radiation Facility using monochromatic radiation
($\lambda$ = 0.3738 \AA). The starting material exhibits an intense
peak at 2$\theta$ = 6.6$^{\circ}$ due to the 002 reflection of the
g-C$_3$N$_4$ TGCN phase (Fig.\ \ref{fig:experimental_data}(A)). During
initial compression to 53 GPa this peak first shifts to larger
2$\theta$ values then broadens and finally disappears above
approximately 40 GPa (Fig.\ \ref{fig:experimental_data}(A)).  This
most likely occurs due to structural disordering associated with layer
buckling and possible interlayer C-N bonding at high pressure as
previously observed for structurally related C$_6$N$_9$H$_3$$\cdot$HCl
\cite{Salamat_pillared-layered_CN_2013}. By 53 GPa the only peaks
remaining in the pattern correspond to the face centered cubic lattice
of solid Ne used as the PTM (Fig.\ \ref{fig:experimental_data}(A)).

Following initial pressurization the material was heated at
1545$\pm$80 K using a CO$_2$ laser beam focused on the sample while
monitoring the X-ray diffraction pattern
\cite{Petitgirard_pyrometry_2014}. After approximately 5 minutes new
single crystal spots appeared in the 2D diffraction pattern that had
been previously dominated by broad diffraction rings from the
compressed TGCN phase and Ne PTM material (Fig.\
\ref{fig:experimental_data}(A) and (B)).  Interestingly, the
appearance of the crystalline peaks was accompanied by a pressure drop
inside the DAC chamber to 44 GPa indicated by ruby fluorescence, as
well as by an 18 \% lattice expansion of the solid Ne used as PTM
(Fig.\ \ref{fig:experimental_data}(A)). This possibly indicates a
corresponding densification within the C$_3$N$_4$ sample. Analysis of
our X-ray data showed no evidence for diamond or any elemental N$_2$
phases formed during the experiment.  There was likewise no evidence
for any of the other predicted carbon nitride phases with different
N/C ratios such as CN or CN$_2$. We conclude that the sample retained
its initial composition close to C$_3$N$_4$ throughout the DAC laser
heating experiment. The crystalline material produced in the
experiment corresponded to only a few crystallites distributed
throughout the sample. While the sample remained at high pressure in
the DAC we carried out a mapping study using a focused ($\sim$2
micrometers) X-ray beam that revealed crystals present in only 3 out
of 36 patterns obtained.  A typical 2D diffraction image is shown in
Fig.\ \ref{fig:experimental_data}(B) showing the crystalline spots
against the sharper rings from the Ne PTM and diffuse scattering from
the densified disordered C$_3$N$_4$ material.

During recovery to ambient conditions the diffraction pattern emerged
more clearly. Once removed from the DAC the sample which remained
mounted inside its Re gasket was examined by X-ray
diffraction. Although there were more crystalline features than at
high pressure the powder averaging remained poor and it was not
possible to carry out \textit{ab initio} or Rietveld structure
refinements ((Fig.\ \ref{fig:experimental_data}(C)).  However we
tested the range of diffraction peaks against calculated patterns for
the various polymorphs produced in our AIRSS study as well as other
theoretically predicted structures. Only one phase contained peaks
within the range of $d$ spacings observed experimentally. This
corresponded to a new C$_3$N$_4$ polymorph with $P4_32_12$ symmetry as
predicted by the AIRSS calculations. This phase is predicted to have a
stability range between 2 to 11 GPa using PBEsol. At 50 GPa it is
highly metastable relative to the predicted $\alpha$-C$_3$N$_4$
($P31c$) or $Pnnm$ structures, but it lies lower in energy than the
metastably compressed graphitic forms (Fig.\
\ref{fig:enthalpy_C3N4_PBE_0-400_GPa}).  We found an excellent match
between all the observed $d$ spacings for the new crystalline phase
and predicted reflections for the $P4_32_12$ polymorph. We then used
Le Bail analysis to obtain refined unit cell parameters $a$ = $b$ =
4.021 \AA, $c$ = 13.922 \AA, $c/a$ = 3.46, and $V$ = 56.26 \AA$^3$ per
f.u., with structural residual or reliability factors R$_{wp}$ = 6.49
and R$_p$ = 4.23 (Fig.\ \ref{fig:experimental_data}(C)). These cell
parameters are in reasonable agreement with the predicted values for
the new structure type at 0 GPa (Table \ref{table:exp_theory}).  The
observed pattern of X-ray intensities differs substantially from those
expected for a fully averaged powder diffraction pattern because only
a few crystallites were present within the laser heated sample. In
addition the crystallites might exhibit preferred orientation further
complicating any more detailed analysis of the experimental data.  We
were able to generate patterns that approximately modelled the
observed intensity distribution by constructing composites containing
a few oriented crystallites with fixed atomic positions from the
$P4_32_12$ theory predictions, applying a March-Dollase function to
correct for preferred orientation. Using this approach we obtained
solutions to a constrained Rietveld refinement of the structure with
large but not unreasonable residual values, e.g., with R$_{wp}$ and
R$_p$ on the order of $\sim$20 (Fig.\ \ref{fig:experimental_data}(D)).

To test the possibility that other atomic arrangements could give rise
to the same unit cell we then used AIRSS to examine alternative
structure solutions within the experimentally determined cell shape
and volume under ambient conditions.  Using the PBEsol functional for
these runs gave a very small residual pressure of about 0.4 GPa in the
experimental cell while PBE indicated a pressure of 3 GPa. Our
searches consistently returned $P4_32_12$ as the most stable phase,
with the next possibility lying at much higher energy (0.75
eV/f.u.). We conclude from this analysis that the newly predicted
$P4_32_12$ polymorph was produced in our high pressure-high
temperature experiment. This is the first recorded experimental
synthesis of a crystalline C$_3$N$_4$ framework structure predicted by
\textit{ab initio} theory. Along with the defective-wurtzite structure
C$_2$N$_3$H
\cite{Horvath-Bordon_high-pressure_CN_2007,Salamat_tetrahedral_CN_2009}
and the recent report of a new CN phase \cite{Goncharov_CN} it
constitutes a new example of a crystalline carbon nitride containing
$sp^3$-bonded carbon atoms.  Various schematic drawings of the new
$P4_32_12$ polymorph are shown in Fig.\
\ref{fig:structure_P4_32_12_Ambientstack} to highlight different
important features of the structure. It corresponds to a new framework
type containing fused triazine rings linked via $sp^3$-bonded C
atoms. This is important because mixed $sp^2/sp^3$ bonding around the
C atoms does not occur for pure carbon polymorphs in equilibrium at
any pressure. In addition the geometrical arrangement containing
$sp^3$-bonded carbon linked to four nitrogen atoms is highly unusual
and perhaps absent in all organic chemistry except perhaps among
transition states.  Our experimental findings indicate that the
probability of forming the new carbon nitride phase is low.  This can
be ascribed to high kinetic barriers as well as the difficulty of
forming the unusual C-N linked tetrahedral geometry associated with
the formation reaction.

\begin{table*}[ht!]
  \caption{\label{table:exp_theory}
    (Top) Comparison of cell parameters for the $P4_32_12$ structure of
    C$_3$N$_4$ at $P=0$ as determined by experiment and theory (LDA, PBEsol and PBE).
    (Bottom) Fractional atomic coordinates from the PBEsol functional. The data in 
    this Table was calculated with a very fine $k$-point
    spacing of 2$\pi\times$0.03\ \AA$^{-1}$.
    %\textcolor{red}{RJN: New fractional coordinates for PBEsol?} 
  }
    \begin{ruledtabular}
%\begin{tabular}{@{\hspace{-1.5cm}}c|c|c|c|c}
\begin{tabular}{c|c|c|c|c}
  Cell parameters     & Experiment & Theory (LDA) & Theory (PBEsol) & Theory (PBE)\\
  \hline
  $a$ (\AA)           &   4.021    &    4.0811     &   4.1106        &    4.1375 \\
  $b$ (\AA)           &   4.021    &    4.0811     &   4.1106        &    4.1375 \\
  $c$ (\AA)           &  13.922    &   14.0490     &  14.1978        &   14.3579 \\
  $c/a$               &   3.46     &    3.442      &   3.454         &    3.470 \\
  $V$/f.u.\ (\AA$^3$) &  56.29      &   58.50      &  59.975          &    61.45 \\
% $V$ (\AA$^3$)       & 225m.15    &  233.991      &  239.898        & 245.792 \\
%  \hline
%  \hline
\end{tabular}
\end{ruledtabular}
\end{table*}

%\vspace{0.25cm}
\begin{table}[ht!]
 \begin{ruledtabular}
\begin{tabular}{c|c|c|c}
  Fractional coordinates & x & y & z  \\
  \hline
  C1     & 0.38982  & 0.20842 & 0.66623  \\
  N2     & 0.24805  & 0.13808 & 0.74384  \\
  N3     & 0.22802  & 0.29442 & 0.58513  \\
  C4     & 0.90820  & 0.09180 & 0.75000  \\
\end{tabular}
\end{ruledtabular}
\end{table}

\section{Structural complexity among carbon nitride polymorphs at low
  pressure}

\begin{figure*}
  \centering
  \includegraphics[width=0.8\textwidth]{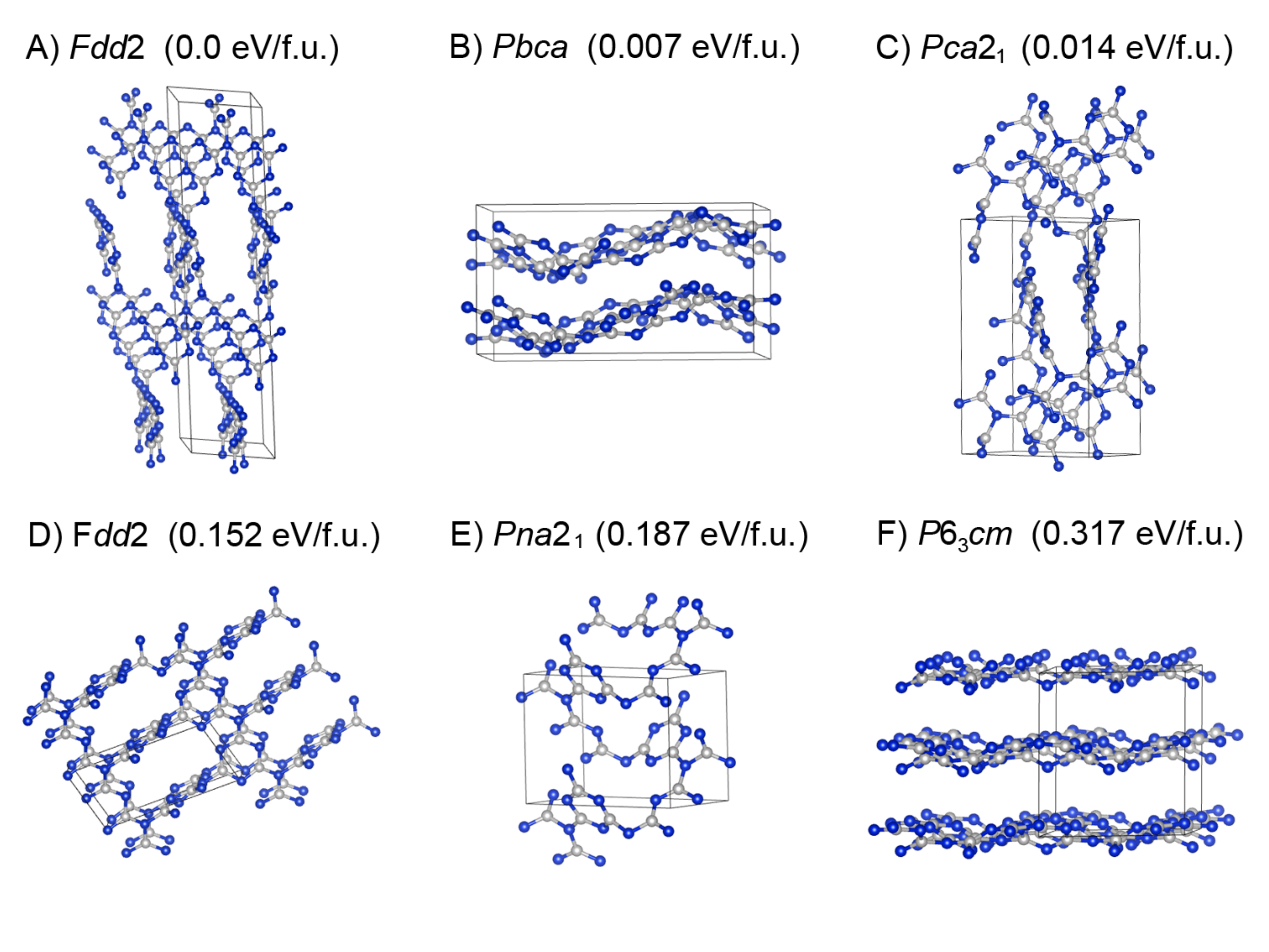}
  \caption{(Color online) Structure types observed among low density
    structures ranked according to relative energy (eV/f.u.) at $P=0$
    GPa using the PBEsol functional.  (A) This ThSi$_2$
    heptazine-based framework with $Fdd2$ symmetry is found to be very
    slightly higher in energy than the ground state PH
    structure of $Cc$ symmetry at zero pressure, according to PBEsol
    calculations \cite{Dong_carbon_nitrides_2015}.  (B) Other
    corrugated layered PH solutions are also competitive as ambient to
    low pressure ground state structures. This $Pbca$ structure lies
    only 0.007 eV/f.u.\ above $Cc$.  (C) Additional three-dimensional
    framework structures based on linked PH units containing
    $sp^2$-bonded atoms are also present at low energy. Here we show
    an example of a $Pca2_1$ symmetry polymorph occurring at 0.014
    eV/f.u.\ above the ground state.  (D) Beginning at 0.152 eV/f.u.\
    above the ground state framework structures based on polytriazine
    imide-linked (PTI) units begin to appear.  (E) The $Pna2_1$
    structure found by AIRSS represents a new example of this PTI
    structure type.  (F) At approximately 0.317 eV/f.u.\ above the
    ground state, PTI corrugated layered structures make their
    appearance.  Here we show the $P6_3cm$ version of the TGCN
    structure used as a starting compound in our laser heated DAC
    experiments.}
  \label{fig:Low_pressure_structures}
\end{figure*}

In the low pressure range where C$_x$N$_y$ compounds are predicted to
be metastable with respect to elemental carbon and nitrogen
\cite{Dong_carbon_nitrides_2015}, it has typically been assumed by
analogy with elemental carbon that layered ``graphitic'' motifs should
constitute the ground state for carbon nitride materials at low
pressures, taking into account the existence of ``voids'' within the
layers due to the different valencies of C and N atoms
\cite{Kroke_Novel_group_14_nitrides_2013,Schwarzer_Tri-s-triazines_2013,Ortega_hexagonal-planar_1995,Lowther_g-CN_1999,Kouvetakis_synthesis_1994,Teter_low_compressibility_1996,Gracia_corrugated_2009,Dong_carbon_nitrides_2015}.
One basic unit of connectivity is the planar triazine (C$_3$N$_3$)
ring that may be interconnected via linking -N= bonds to form
polytriazine imide (PTI) layers with C$_3$N$_4$ stoichiometry
\cite{Ortega_hexagonal-planar_1995,Lowther_g-CN_1999,Kouvetakis_synthesis_1994,Teter_low_compressibility_1996,Algara-Siller_triazine_g-CN_2014}. The
polymeric TGCN material synthesized by Algara-Siller \textit{et al.}\
\cite{Algara-Siller_triazine_g-CN_2014} and used as the precursor for
our high pressure experiments was based on such PTI layers with either
AB (space group $P\bar{6}m2$) or ABC ($P6_3cm$) stacking. The most
stable TGCN polymorph found in our AIRSS study had $P6_3cm$ symmetry
(Fig.\ \ref{fig:Low_pressure_structures}). It is predicted to lie at
0.329 eV/f.u.\ above the ground state at ambient pressure (see
Supplemental Table I). It has been shown that layers based on linked
tri-s-triazine (polyheptazine:\ PH) units are more stable than the PTI
motifs
\cite{Kroke_Novel_group_14_nitrides_2013,Gracia_corrugated_2009}. Various
levels of complexity can appear among both PTI and PH connectivity
patterns within a single layer, and the layers are predicted to be far
from planar. Substantial buckling around the bridging -N= units that
act as mechanical hinges lead to structural models with ``corrugated''
layers \cite{Gracia_corrugated_2009}.

\begin{figure*}
  \includegraphics[width=5.9cm]{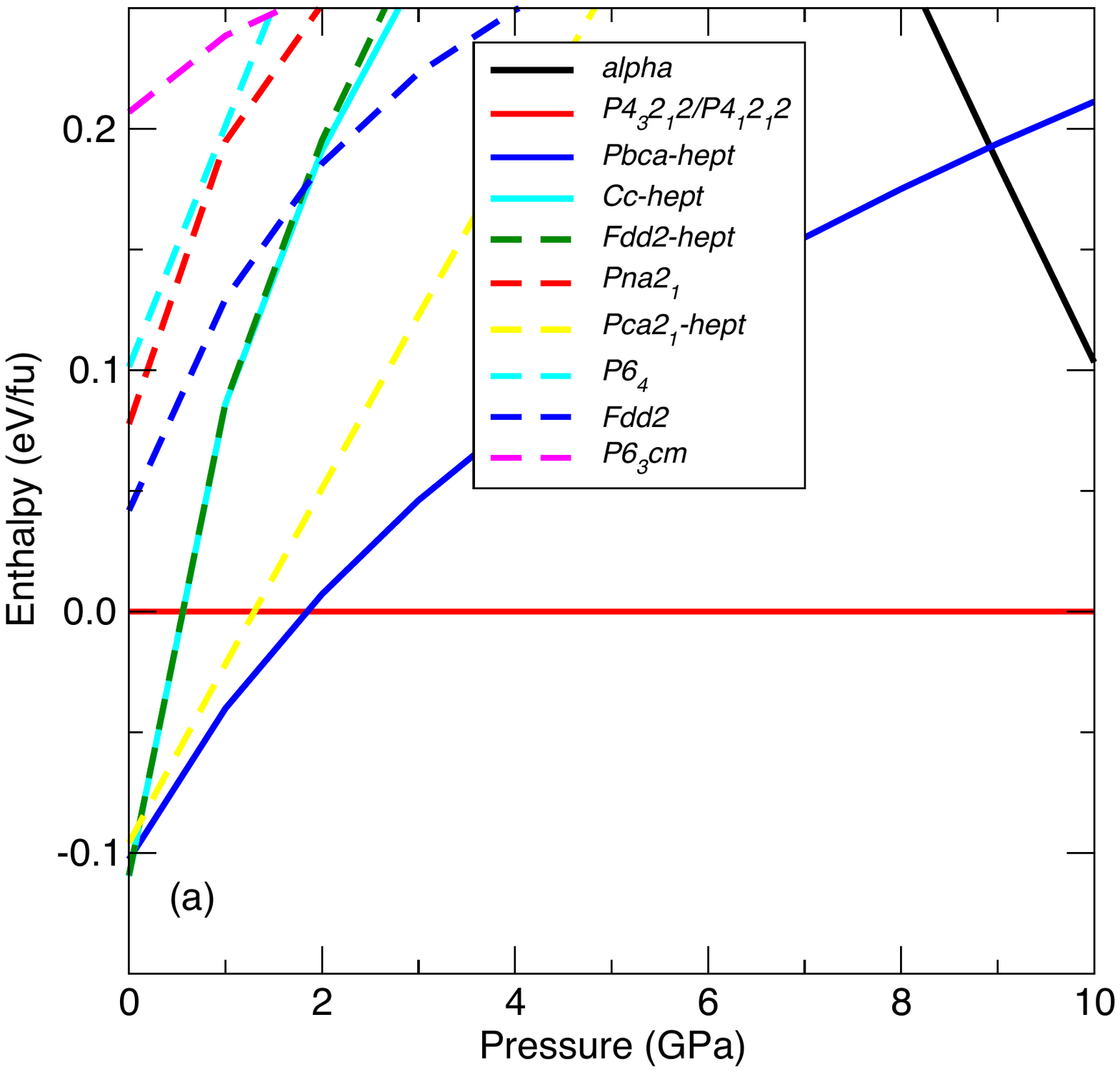}
  \includegraphics[width=5.9cm]{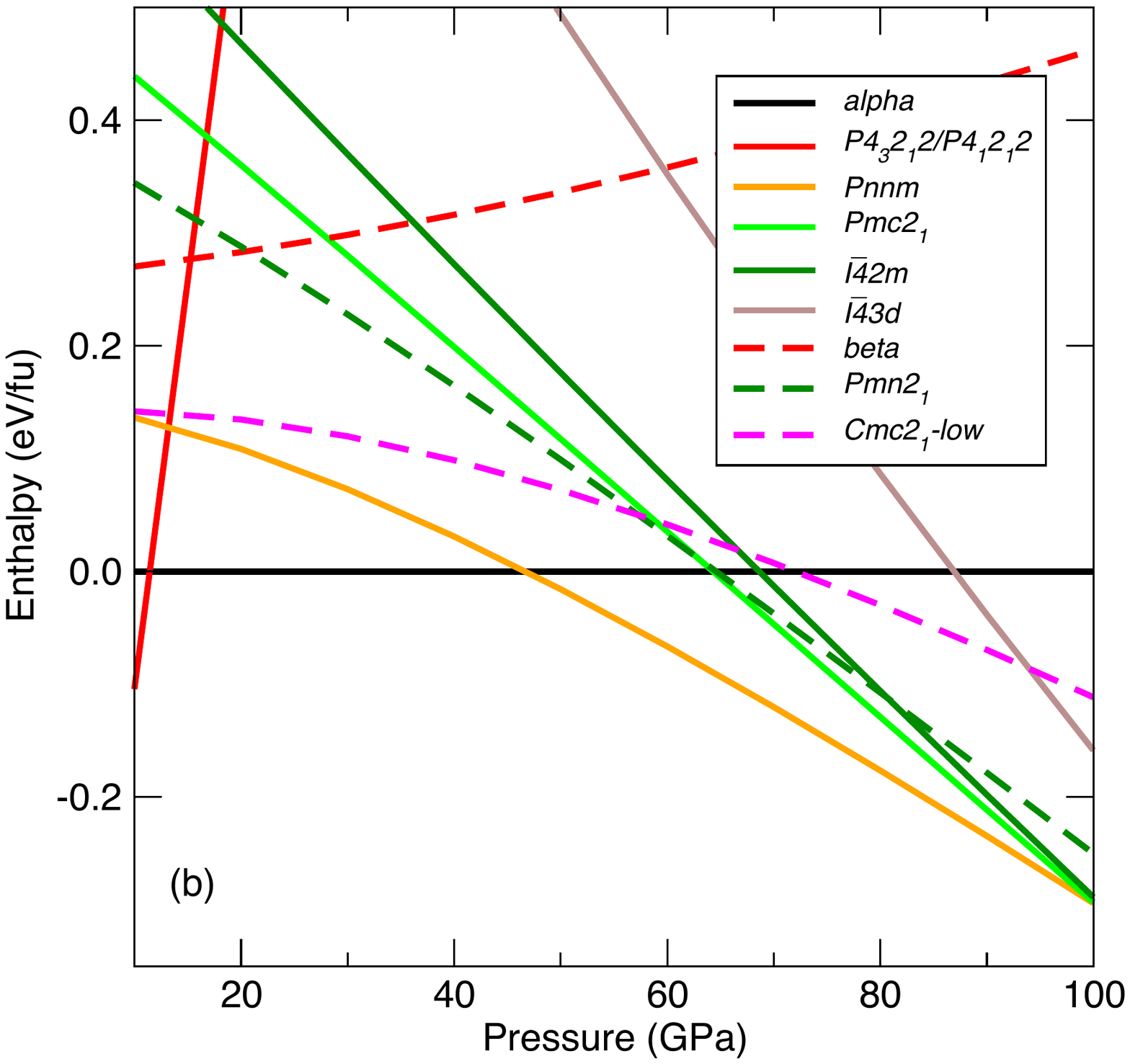}
  \includegraphics[width=5.9cm]{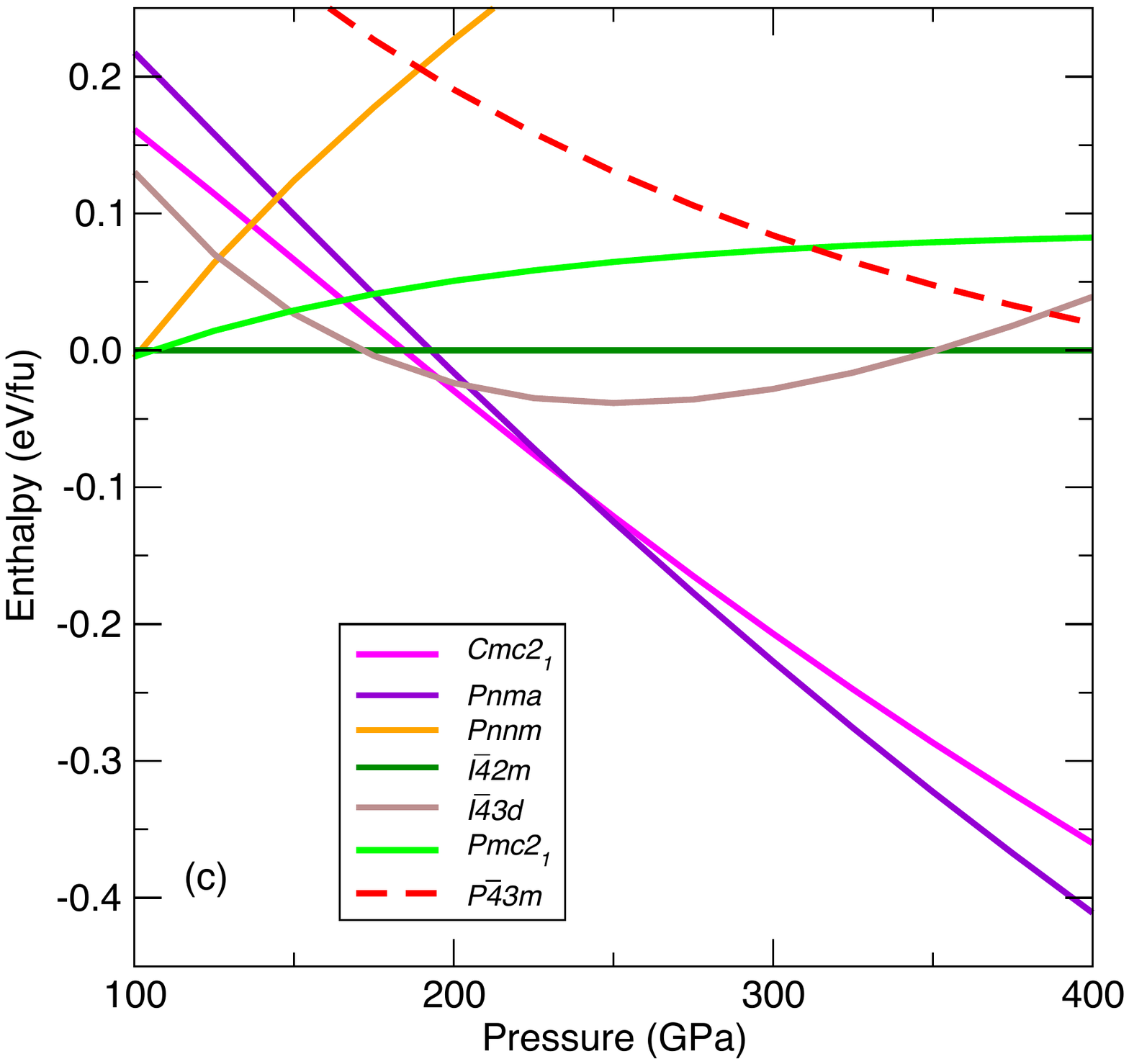}
  \caption{(Color online) PBE Enthalpy versus pressure in C$_3$N$_4$
    in pressures up to 400 GPa.  Enthalpy per atom using the PBEsol
    functional as a function of pressure for C$_3$N$_4$ polymorphs
    identified by AIRSS.  The solid lines indicate
      phases that are predicted to have stable regions in the range
      shown.  Results are shown in three pressure ranges to highlight
    polymorphism at low (0--10 GPa), intermediate (10--100 GPa) and
    high (100--400 GPa) pressures. In (a) the enthalpies are plotted
    relative to the $P4_32_12$ framework polymorph, whereas in (b) the
    $\alpha$-C$_3$N$_4$ structure is used as a reference state. In (c)
    the $I\bar{4}2m$ structure is taken as the reference.}
  \label{fig:enthalpy_C3N4_PBE_0-400_GPa}
\end{figure*}

Dong \textit{et al.}\
  \cite{Dong_carbon_nitrides_2015} found a $Cc$ structure to be the
  ground state at 0GPa, and  we
  have found a very similar structure of $Fdd2$ symmetry that is only
  a fraction of an meV per f.u.\ higher in energy than $Cc$ (see Fig.\
  \ref{fig:Low_pressure_structures} and Supplemental Material
  \cite{Supplemental}).  We have also found a new
  corrugated PH structure of $Pbca$ symmetry that is predicted to be
  about 7 meV per f.u.\ higher in energy than $Cc$ andon analysis is the same
  as a structure found by Gracia and Kroll but in a larger unit cell
  with $P1$ symmetry.  We
  predict that $Pbca$ is the most stable phase in the range 0.05--2
  GPa.  The $Cc$, $Fdd2$, and $Pbca$
  structures are based on the ThSi$_2$ motif
  \cite{Gracia_corrugated_2009}.  The $Pbca$ phase, which has a starting
  volume near 73 \AA$^3$/f.u.\, is expected to cross the $P6_3cm$
  enthalpy-pressure curve at approximately 25 GPa and so would not be
  accessible to the experimental synthesis procedure (Fig.\
  \ref{fig:enthalpy_C3N4_PBE_0-400_GPa}).  Gracia and Kroll
  \cite{Gracia_corrugated_2009} noted tubular PH structures that could
  lead to macroscopic forms analogous to carbon nanotubes.

\begin{figure*}
\includegraphics[width=15cm]{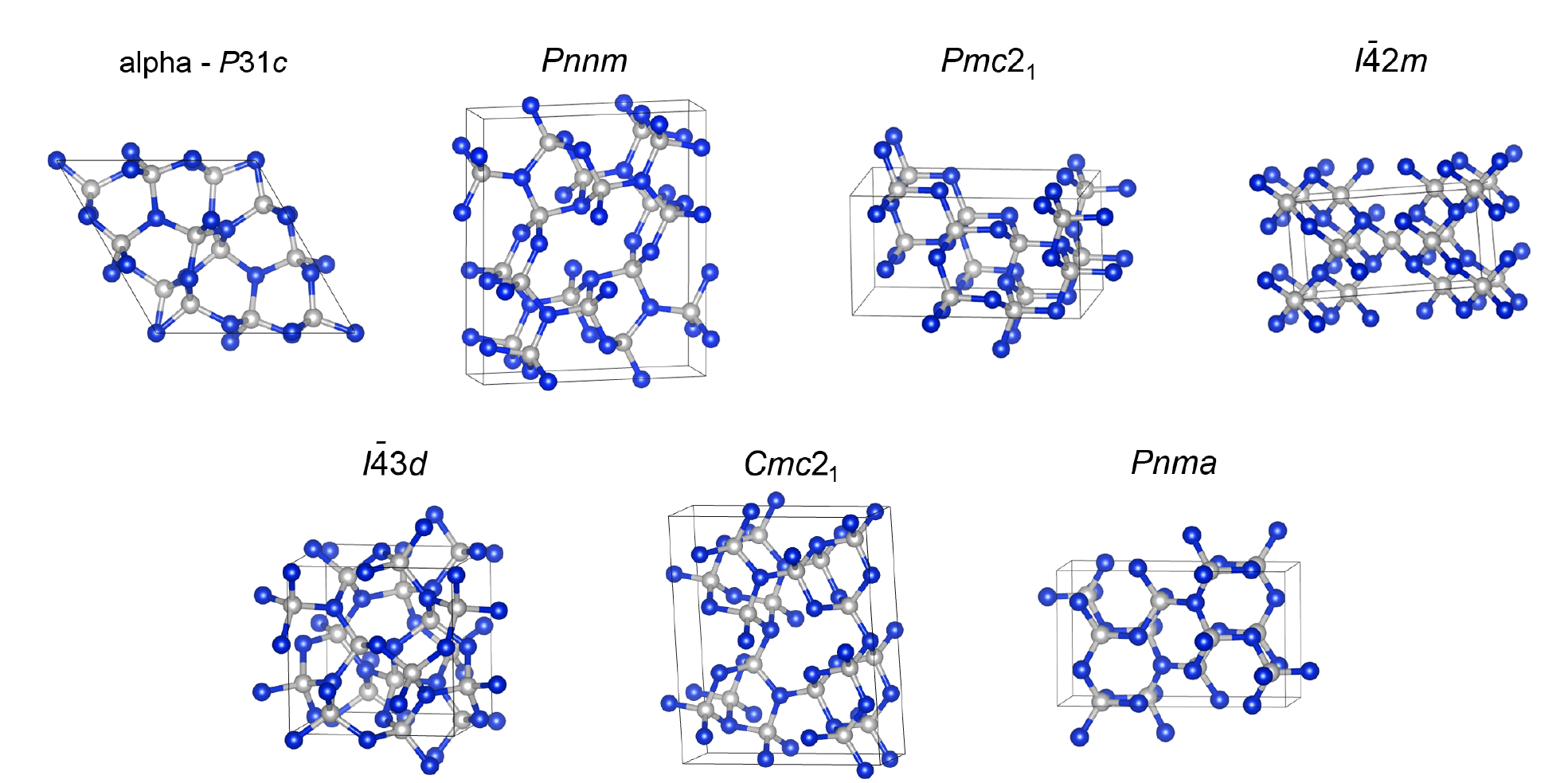}
\caption{(Color online) Structures formed at intermediate and high
  pressure in the C$_3$N$_4$ system.  N atoms are depicted as blue
  spheres and C atoms are white/grey. The $\alpha$-C$_3$N$_4$ ($P31c$)
  phase is predicted to be stable between 11--47 GPa using PBEsol,
  followed by $Pnnm$, $Pmc2_1$, $I\bar{4}2m$, $I\bar{4}3d$, $Cmc2_1$
  and $Pnma$ polymorphs at the transition pressures given in the
  text.}
  \label{fig:structures_high_pressure}
\end{figure*}

\begin{figure}
  \centering
  \includegraphics[width=0.4\textwidth]{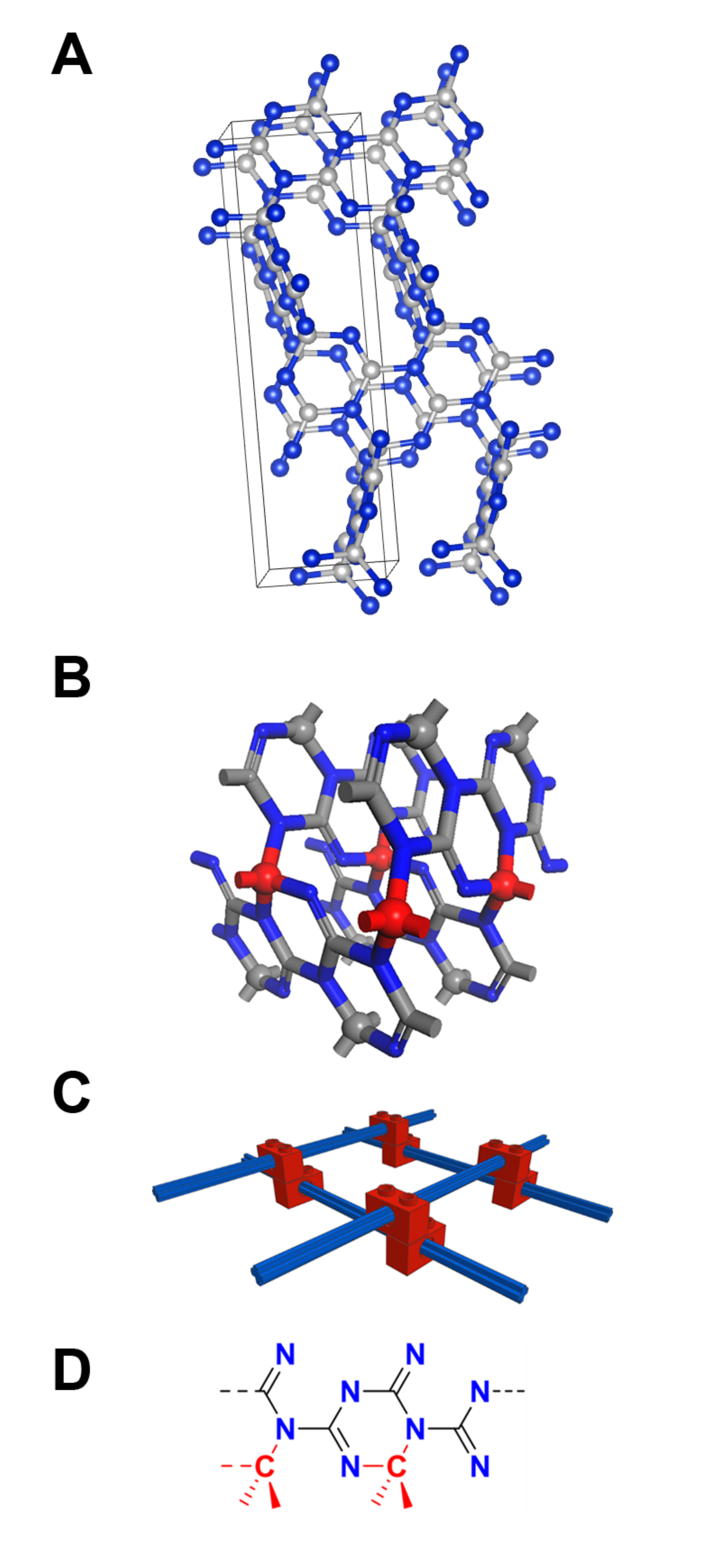}
  \caption{(Color online) Different ways of depicting the $P4_32_12$
    structure.  (A) Ball and stick model: carbon atoms - grey,
    nitrogen atoms - blue. (B) Structure drawing emphasizing the
    $sp^3$-bonded carbon atoms (red). (C) Abstract ``Lego''
    representation featuring parallel, 1D struts of quasi-aromatic
    imidamides (shown in blue) linked together by tetrahedral
    $sp^3$-bonded carbon joints (shown in red). (D) A fragment of the
    hybrid $sp^2$-$sp^3$ bonded carbon nitride.}
  \label{fig:structure_P4_32_12_Ambientstack}
\end{figure}

Many of these structure types have very large unit cell volumes
($>$100 \AA$^3$/f.u.) that are rapidly destabilized at high pressure,
see Fig.\ \ref{fig:enthalpy_C3N4_PBE_0-400_GPa}. Others with energies
up to 1 eV/f.u.\ above the ground state have volumes in the 65--90
\AA$^3$/f.u.\ range and they appear as metastable structure solutions
in Fig.\ \ref{fig:structures_high_pressure}. One structure stands
out. This is the new $P4_32_12$ phase described above that we have
synthesized experimentally (Fig.\
\ref{fig:structure_P4_32_12_Ambientstack}).  The experimental volume
of $P4_32_12$ of 56.26 \AA$^3$/f.u.\ is a little smaller than the
PBEsol volume of 59.97 \AA$^3$/f.u.

A further new phase of $Fdd2$ symmetry consisting of triazine rings
linked into a three-dimensional framework via $sp^2$-bonded atoms is
found to occur at 0.152 eV/f.u.\ above the ground state at ambient
pressure (Fig.\ \ref{fig:Low_pressure_structures}). The identification
of such a new family of framework structures built from thermally
stable C-N bonds opens up new directions for carbon nitride
research. The challenge for carbon nitride synthesis is now to produce
some of these predicted phases and their unusual bonded architectures
in a controlled way, and verify their structures and determine their
properties experimentally.

Of particular interest is the behavior of the $P6_3cm$ phase that
forms the main constituent of TGCN used as a starting material for our
laser heated diamond anvil cell experiments. At ambient pressure it
lies at 0.317 eV/f.u.\ above the ground state and at approximately 0.2
eV/f.u.\ above $P4_32_12$, see Fig.\
\ref{fig:enthalpy_C3N4_PBE_0-400_GPa}.  During initial compression the
TGCN phase loses its initial crystallinity but it can be expected to
continue along a similar enthalpy-pressure path to reach approximately
0.5 eV/f.u.\ above the $P4_32_12$ energy by 50 GPa. By this pressure
$P4_32_12$ has become substantially metastable (by $\sim$3 eV/f.u.)
relative to $\alpha$-C$_3$N$_4$ and other dense phases but it is the
next lowest in enthalpy that we have found following
destabilization of the disordered and metastably compressed $P6_3cm$
material.

These predicted results can allow us to better understand the
nucleation of only a few crystals of the $P4_32_12$ phase following
laser heating within the metastably compressed $P6_3cm$ TGCN
compound. Interestingly, several other C$_3$N$_4$ polymorphs are also
predicted to occur within the 20--80 GPa pressure range, with
enthalpies extending to $\sim$0.5 eV/f.u.\ above the ground state. We
might expect that further high-$P$,$T$ experiments carried out on
metastably compressed starting materials could lead to identification
of these new C$_3$N$_4$ polymorphs.

%\vspace{0.25cm}

%Table 1 : Structural parameters for carbon graphite and diamond and
%the graphite-diamond transition pressure ($T=0$ K) calculated using
%five different density functionals (LDA, PBE, PBEsol, PBE-G06 and WC).

%\vspace{0.25cm}

\section{Conclusions}

Extended $sp^2$-bonded framework and layered carbon nitride structures
are found to be thermodynamically stable at pressures
below about 2 GPa.  The lowest energy structures are built from linked
polyheptazine (PH) units: at higher energies layered and 3D
microporous frameworks based on triazine rings linked by imide groups
(PTI structures) appear. The enthalpies and volumes per f.u.\ are
reported in the Supplemental Material \cite{Supplemental}.

We find that the microporous PH $Cc$ structure is the most stable at
zero pressure, although the similar $Fdd2$ structure is only a
fraction of an meV per f.u.\ higher in energy. A layered $Pbca$ PH structure
is predicted to be stable in the range 0.05--2 GPa.

The $P4_32_12$ structure is a new structure prototype
  not found in databases that we have discovered using AIRSS. It has
  mixed $sp^2$ and $sp^3$ bonding and is predicted to
be stable in the range 2--11 GPa. C$_3$N$_4$ therefore differs
from pure carbon in that there is a thermodynamically stable phase
between the $sp^2$ and $sp^3$ bonded forms.  This structure could be
accessible through suitable precursors and synthetic routes at low
pressure.

All of the predicted thermodynamically stable phases at pressures
above 11 GPa consist entirely of four-fold coordinated C atoms and
three-fold coordinated N atoms.

The relative stabilities of the most favorable C$_3$N$_4$ have
uncertainties due to the use of approximate density functionals which
amount to a few GPa at low pressures.  We have, however, sought to
assess the sizes of these errors by comparing with results for the
related graphite/carbon diamond system.

AIRSS has predicted new $Pbca$, $P4_32_12$, $Pnnm$, $Pmc2_1$,
  and $Pnma$ C$_3$N$_4$ structures to have regions of thermodynamic
  stability and, in addition, the new $Fdd2$ structure is almost
  stable.  Our calculations predict decomposition of C$_3$N$_4$ at
pressures above 650 GPa into diamond and pyrite-structured CN$_2$.

Our experimental study has provided strong evidence for the synthesis
in laser-heated diamond anvil cell experiments of the newly-predicted
$P4_32_12$ structure.  The identification of a new family of
structures with mixed $sp^2$ and $sp^3$ bonding opens up a new area of
exploration for these important photo- and catalytically active solid
materials. 

\newpage

\begin{acknowledgments}

  R.\ J.\ N.\ acknowledges financial support from the Engineering and
  Physical Sciences Research Council (EPSRC) of the U.K.\
  [EP/J017639/1].  C.\ J.\ P.\ acknowledges financial support from
  EPSRC [EP/G007489/2] and [EP/K013688/1], and a
  Royal Society Wolfson Research Merit Award​.  R.\ J.\ N.\ and C.\ J.\
  P.\ acknowledge use of the Archer facilities of the U.K.'s national
  high-performance computing service (for which access was obtained
  via the UKCP consortium [EP/K013688/1] and EP/K014560/1).  P.\ F.\
  M.\ was supported by EPSRC grant [EP/L01709/1].

\end{acknowledgments}

\end{document}